\documentclass[showpacs,amsmath,amssymb,prb,onecolumn]{revtex4}
\usepackage{graphicx}
\usepackage{dcolumn}
\usepackage{bm}

\begin{document}
\title{Effects of different electron-phonon couplings on spectral and transport properties of small molecule single-crystal organic semiconductors}

\author{C. A. Perroni $^{1,}$*, F. Gargiulo $^{2}$, A. Nocera $^{3}$, V. Marigliano Ramaglia $^{1}$ and V. Cataudella $^{1}$}

\affiliation{
$^{1}$ CNR-SPIN and Dipartimento di Fisica, Univ. ``Federico
II'', Via Cinthia, Napoli, I-80126, Italy \\
$^{2}$ Institute of Theoretical Physics, \'{E}cole Polytechnique F\'{e}d\'{e}rale de Lausanne (EPFL), CH-1015 Lausanne, Switzerland \\
$^{3}$ Department of Physics, Northeastern University - Boston, MA 02115, USA}

\begin{abstract}
Spectral and transport properties of small molecule single-crystal organic semiconductors have been theoretically analyzed focusing on oligoacenes, in particular on the series from naphthalene to rubrene and pentacene aiming to show that the inclusion of different electron-phonon couplings is of paramount importance to interpret accurately the properties of prototype organic semiconductors.
While, in the case of rubrene, the coupling between charge carriers and low frequency inter-molecular modes is sufficient for a satisfactory  description of spectral and transport properties, the inclusion of electron coupling to both low frequency inter-molecular and high frequency intra-molecular vibrational modes is needed to account for the temperature dependence of transport properties in smaller oligoacenes. \\
For rubrene, a very accurate analysis in the relevant experimental configuration has allowed to clarify the origin of the temperature dependent mobility observed in these organic semiconductors. With increasing temperature, the chemical potential moves into the tail of the density of states corresponding to localized states, but this is not enough to drive the system into an insulating state. The mobility along different crystallographic directions has been calculated, including vertex corrections that give rise to a transport lifetime one order of magnitude smaller than the spectral lifetime of the states involved in the transport mechanism. The mobility always exhibits a power-law behavior as a function of temperature in agreement with experiments in rubrene.\\
In systems gated with polarizable dielectrics, the electron coupling to interface vibrational modes of the gate has to be included in addition to the intrinsic electron-phonon interaction. While the intrinsic bulk electron-phonon interaction affects the behavior of mobility in the coherent regime below room temperature, the coupling with interface modes is dominant for the activated high temperature contribution of localized polarons.\\
Finally, the effects of a weak disorder largely increase the activation energies of mobility and induce the small polaron formation at lower values of electron-phonon couplings in the experimentally relevant temperature window.
\end{abstract}

\maketitle


\section{Introduction}

In recent years, the interest in plastic electronics has grown considerably. The realization of devices such as organic field-effect transistors (OFET) represents a key step in this field. Single-crystal OFET made of ultrapure small molecule semiconductors are characterized by mobilities up to one order of magnitude larger then those typical of thin film transistors \cite{takeya}. The most promising are those based on oligoacenes, such as pentacene and rubrene, which exhibit a strong anysotropy and the largest mobility measured in organic semiconductors \cite{morpurgo}.

In spite of many applications based on such devices, the intrinsic transport mechanism acting in high mobility organic semiconductors is not fully understood. Transport measurements from $100$ K to room temperature in single crystal semiconductors, such as rubrene, show a behaviour of the charge carrier mobility $\mu$ which can be defined band-like ($\mu\propto T^{-\gamma}$, with the exponent $\gamma$ close to two) similar to that observed in crystalline inorganic semiconductors \cite{morpurgo}.
However, the order of magnitude of mobility is much smaller than that of pure inorganic semiconductors, and the mean free path for the carriers
has been theoretically estimated to be comparable with the molecular separation at room temperature \cite{Cheng}. Therefore, the Ioffe-Regel limit is reached with increasing temperature. Moreover, in some systems, starting from room temperature, a crossover from band-like to activated hopping behavior can take place \cite{corop,cheng1,hanne3}. The crossover has been interpreted as due to the formation of the polaron, that is the quasi-particle formed by the electron (or hole) and the surrounding phonon cloud \cite{alex}. For example, in naphthalene and anthracene, while the mobility along the $a$ and $b$ axis shows only a slightly change with the temperature, that along the $c$ axis is characterized by a temperature activated behavior at higher temperature with energy barrier of the order of $15$ meV \cite{warta}. The experimental data in these compounds suggest that the coherent band transport is gradually destroyed and the transport due to polaron hopping evolves as a parallel channel dominating at sufficiently high temperature (which can be larger than room temperature) \cite{warta}.

\begin{figure}[htb]
\centering
\includegraphics[width=0.75\textwidth]{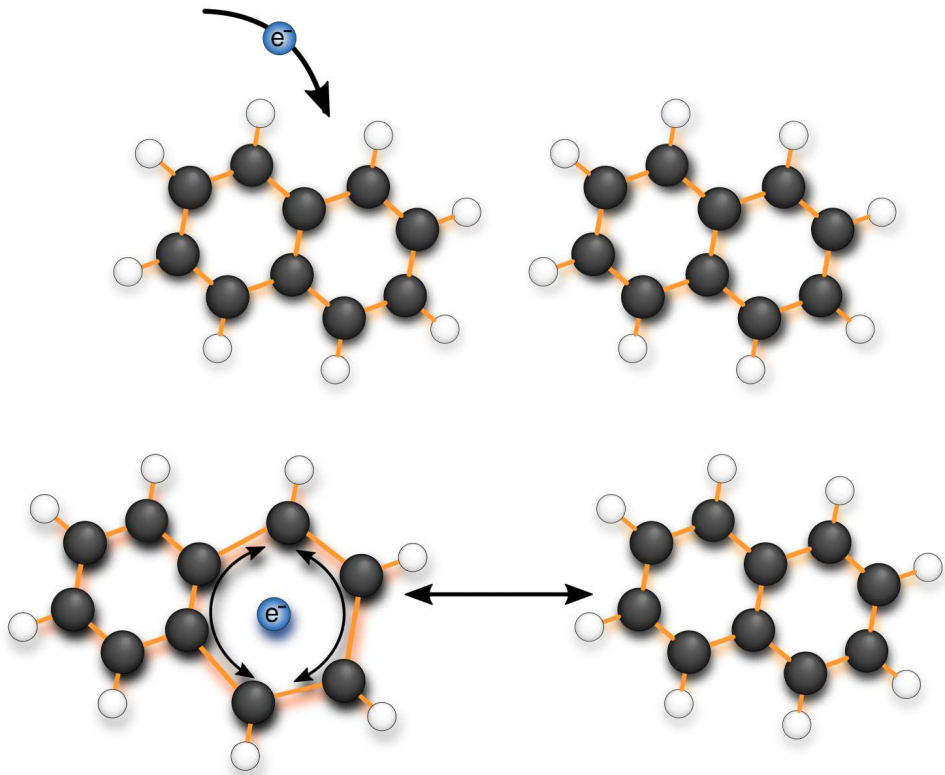}
\caption{\label{naphta} Sketch about the interplay between low frequency inter-molecular and high frequency intra-molecular vibrational modes in a crystal of naphtalene. The charge carrier deforms the benzene ring when it is on the molecule. Moreover, the charge carrier displaces two neighbor molecules when it jumps.}
\end{figure}

In systems with polarizable gates, scaling laws of the mobility as a function of the dielectric constant of solid \cite{stassen,nature} and liquid \cite{ono} gates have been discovered pointing out that the nearby dielectric has a strong influence. Actually, if the difference between the dielectric constant of the organic semiconductor and of the gate is small, at temperatures close or higher than $100$ K, the mobility $\mu$ of these systems exhibits the power-law band-like behavior.  On the other hand, if the dielectric constant mismatch is high, an activated insulating behavior is found with much smaller values of mobility at room temperature \cite{nature}. A possible explanation of this behavior is that the injected charge carriers undergo a polaronic localization due to the interaction with modes at the interface with the polarizable dielectric gate \cite{nature,fratini}.

Extended vs. localized features of charge carriers appear also in spectroscopic observations. Angle resolved photoemission spectroscopy
(ARPES) supports the extended character of states \cite{Ostrogota,Ostrogota1,arpes2} showing that the quasi-particle
energy dispersion does exhibit a weak mass renormalization even if the width of the peaks of the spectral function increases significantly with temperature. For pentacene, the bandwidth is reduced only by about $15 \%$ going from $75 K$ to $300 K$ indicating moderate values of electron-phonon coupling.
On the other hand, some spectroscopic probes, such as electron spin resonance (ESR) \cite{Marumoto1,Marumoto2}, THz \cite{Laarhoven},
and modulated spectroscopy \cite{Sakanoue} are in favor of states localized within few molecules. Actually, in rubrene and in pentacene, to ascribe the presence of localized features to small polarons is not likely since the electron-phonon coupling is not large enough to justify the polaron formation \cite{corop}. Therefore, one of the main theoretical problems is to conciliate band-like with localized features of charge carriers \cite{Troisi Orlandi}.

First-principle calculations have pointed out that charge carriers are affected by the coupling to inter-molecular modes with low frequency in comparison with typical electron hoppings \cite{corop,bredas}.  A model that is to some extent close to the Su-Schrieffer-Heeger (SSH) \cite{SSH} hamiltonian has been recently introduced to take into account this interaction \cite{Troisi Orlandi}.
It is a minimal one-dimensional (1D) system, valid for the most conductive crystal axis of high mobility systems,  where the effect of the electron-phonon coupling is reduced to a modulation of the transfer integral \cite{Troisi Rubrene}.
A dynamic approach where vibrational modes are treated as classical variables has been used in 1D and in a recent generalization to two dimensions (2D) \cite{Troisi2D}. Within this method, the temperature dependence of computed mobility is in agreement with experimental
results. However, the role of dimensionality of the system is not clear: in fact, in the 1D case, one has $\mu\propto T^{-2}$, while,
in the 2D case, the decrease of the mobility with temperature is intermediate between $\mu\propto T^{-2}$ and $\mu\propto T^{-1}$. In any case, the computed mobility is larger than that measured (at least a factor of two). Moreover, the dynamics of only one charge particle is studied neglecting completely the role of the chemical potential. Finally, the effects on charge carrier dynamics due to the coupling with vibrational modes are included in an approximate way \cite{licorop} and the corresponding coupled dynamics do not recover the right thermal equilibrium on long times.

Recently, the transport properties of the 1D SSH model have been analyzed within a different adiabatic approach \cite{Ciuchi Fratini} mapping the problem onto that of a single quantum particle in a random potential (generalized Anderson problem \cite{Anderson}). Very recently, some of us have made a systematic study of this 1D model including the vertex corrections into the calculation of the mobility \cite{vittoriocheck}. While finite frequency quantities are properly calculated in this 1D model, the inclusion of vertex corrections leads to a vanishing mobility unless an ad-hoc broadening of the energy eigenvalues is assumed.

It is clear that 1D adiabatic models suffer of severe limitations, of which the main is that electronic states are always localized \cite{Anderson}. Moreover, features such as band anisotropy, small but finite carrier density are necessary for a correct description of the systems. Therefore, in this review, we first analyze a generic three-dimensional (3D) model such as the anisotropic Holstein model \cite{hannewald} in order to discuss the relevant issue of the band anisotropy at finite carrier density. This model is studied within the adiabatic approach focusing on the weak to intermediate el-ph coupling regime which is relevant for high mobility organic semiconductors \cite{meholstein}. Next, we analyze a realistic model for rubrene which represents an extension of the 1D SSH model to the quasi 2D case since this is the relevant geometry for OFET \cite{fernando}.

Spectral and transport properties calculated within these two models are discussed in this review. The spectral functions show peaks which are weakly renormalized in comparison with those of the bare bands. However, with increasing temperature, the width of the spectral functions gets larger and larger making the quasi-particles less defined. The marked width of the spectral functions gives rise to densities of states with a low energy exponential tail
increasing with temperature. At low temperatures, this tail corresponds to localized states and gives rough indications for the energy position of the mobility edge. With increasing temperature, in the regime of low carrier doping appropriate to most OFET, the chemical potential always enters the energy region of the tail. The features of the spectral function and the behavior of the chemical potential allow to reconcile the band-like description (ARPES data) with the finding that charge carriers appear more localized at high temperature (ESR and modulated spectroscopy data). The study of spectral properties also clarifies that the states that mainly contribute to the conduction process have low momentum and are not at the chemical potential. The mobility $\mu$ is studied as a function of the electron-phonon coupling, the temperature  and particle density.
Not only the order of magnitude and the anisotropy ratio between different directions are in agreement with experimental observations, but also the temperature dependence of $\mu$ is correctly reproduced in the model for rubrene since it scales as a power law $T^{-\gamma}$, with $\gamma$ close or larger than two. The inclusion of vertex corrections in the calculation of the mobility is relevant, in particular, to get a transport lifetime one
order smaller than the spectral lifetime of the states involved in the transport mechanism. Moreover, with increasing temperature, the Ioffe-Regel limit is reached since the contribution of itinerant states to the conduction becomes less and less relevant.

{\it Ab-initio} calculations have clarified that charge carriers in organic semiconductors are not only coupled to low frequency inter-molecular modes, but also to intra-molecular modes with high frequency in comparison with typical electron hoppings \cite{corop,hanne2} (see Fig. \ref{naphta} for a sketch in naphtalene crystal). An important point is that the reorganization energy (related to the polaron binding energy) decreases with increasing the number of benzene rings in oligoacenes (for example, going from naphthalene to pentacene). In order to fully explore the effects of the different modes on prototype single crystal organic semiconductors, such as oligoacenes, a model with intermediate coupling to both intra- and inter-molecular modes is analyzed in this review \cite{meholpssh}. We will show that the interplay between local and non local electron-phonon interactions is able to provide a very accurate description of the mobility and to shed light on the intricate mechanism of band narrowing with increasing temperature \cite{arpes2}.

When the organic semiconductor is grown on a polarizable gate, it is important to analyze the effects of electron coupling to surface vibrational modes of the gate at the interface with the semiconductor mediated by a long-range electron-phonon interaction  \cite{substrato,mebipo}. In this review, we analyze a model which combines the effects of interface and intrinsic bulk electron-phonon couplings on the transport properties at finite temperature. We show that the coupling to the organic semiconductor bulk phonon modes affects the behavior of mobility below room temperature enhancing the coherent contribution, but it is ineffective on the incoherent small polaron contribution dominated by the interface coupling at high temperatures.

In order to improve the modeling of organic semiconductors, the effect of a weak disorder due to bulk and interface traps is included \cite{substrato}. In particular, the interplay between long-range electron-phonon interactions and disorder effects is investigated within a model. The disorder effects are able to enhance the hopping barriers of the activated mobility and to drive the small polaron formation to lower values of electron-phonon interactions. We point out that disorder is a key factor to get agreement with experimental data in rubrene OFET grown on polarizable gate dielectrics, such as the $Ta_2 O_5$ oxide \cite{nature}.

The paper is organized in the following way. In section II, the effects of electron coupling to low frequency vibrational modes on the spectral and transport properties are discussed in high-dimensional Holstein-like and SSH-like models. In section III, the effects of electron coupling to both low frequency inter-molecular and high frequency intra-molecular modes on the spectral and transport properties are investigated. In section IV, the influence of gates made of polarizable dielectrics and the interplay between electron-phonon couplings and disorder strength on the transport properties are emphasized. In section V, conclusions and final discussions.


\section{Effects of low frequency vibrational modes}

The anisotropy of the electronic properties is a key ingredient in the description of organic semiconductors \cite{hannewald}. Therefore, in subsection $2.1$, we will introduce a simple anisotropic tight-binding model including a Holstein-like electron coupling to low frequency modes in order to focus on the effects of electronic structure.  However, the results will be discussed later in subsection $2.4$ in comparison with those obtained by a more detailed model based on inter-molecular low frequency modes that will be introduced in subsection $2.2$.

\subsection{Band anisotropy}

We assume the following model hamiltonian:
\begin{equation}
H= - \sum_{\vec{R}_i, \vec{\delta}} t_{\vec{\delta}} c_{\vec{R}_i}^{\dagger}c_{\vec{R}_i+\vec{\delta}}+
\sum_{\vec{R}_i} \frac{{p}^2_{\vec{R}_i}}{2 m}+\sum_{\vec{R}_i} \frac{ k x_{\vec{R}_i}^{2}}{2}+H_{el-ph},
\label{h}
\end{equation}
where $t_{\vec{\delta}}$ is the bare electron hopping toward the nearest neighbors $\vec{\delta}$, $c_{\vec{R}_i}^{\dagger}$ and $c_{\vec{R}_i}$ are the charge carrier creation and annihilation operators, respectively, relative to the site $\vec{R}_i$ (of a cubic lattice with parameter $a$ for the Holstein model and of an orthorombic lattice with constants $a$, $b$, $c$ for the rubrene model of the next section), $x_{\vec{R}_i}$ and ${p}_{\vec{R}_i}$ are the oscillator displacement and momentum, respectively, $m$ the oscillator mass, and $k$ the elastic constant. In eq.(\ref{h}), $H_{el-ph}$ represents the electron-phonon coupling term.

A simple model able to capture the anisotropy of the electronic properties of these materials is the anisotropic Holstein hamiltonian which is generic for high mobility organic semiconductors such as oligoacenes \cite{meholstein}. This model assumes a cubic lattice with the following anisotropic hopping integrals in Eq. (\ref{h}):
$t_z \simeq$ 100 meV, $t_x \simeq 50$ meV, $t_y \simeq 20$ meV. Moreover, typical values of phonon frequency $\omega_0 =\sqrt{k/m}$ are of the order of $10 meV$ leading to a very low adiabatic ratio $\gamma=\omega_0/t_z = 0.1$ (adiabatic regime) \cite{hannewald}. Finally, we assume a very general electron-phonon interaction inspired by Holstein model \cite{hannewald,holstein1,holstein2}. Therefore, in Eq.(\ref{h}), the electron-phonon hamiltonian is
\begin{equation}
H_{el-ph}=\alpha \sum_{\vec{R}_i} x_{\vec{R}_i} c_{\vec{R}_i}^{\dagger} c_{\vec{R}_i},
\label{hel}
\end{equation}
where $\alpha$ is the coupling constant controlling the link between the local electron density and lattice displacement. The following dimensionless quantity $\lambda_{Hol}$
\begin{equation}
\lambda_{Hol}=\frac{\alpha^2}{4 k t_z} \label{lamb}
\end{equation}
correctly describes the strength of the electron-phonon coupling.

\subsection{Inter-molecular vibrational modes}

In this section, we analyze the influence of the electron coupling to low-frequency inter-molecular modes on the properties of small molecule organic semiconductors.

We consider a realistic quasi-2D model which simulates the properties of rubrene \cite{fernando}. Therefore, it not only includes the anisotropy of rubrene crystals (shared with many other small molecule organic semiconductors), but also a more appropriate electron-phonon coupling. First, we derive the effective low-energy electronic model, then the lattice parameters and the appropriate electron-phonon interaction.

\begin{figure}[htb]
\centering
\includegraphics[width=0.75\textwidth]{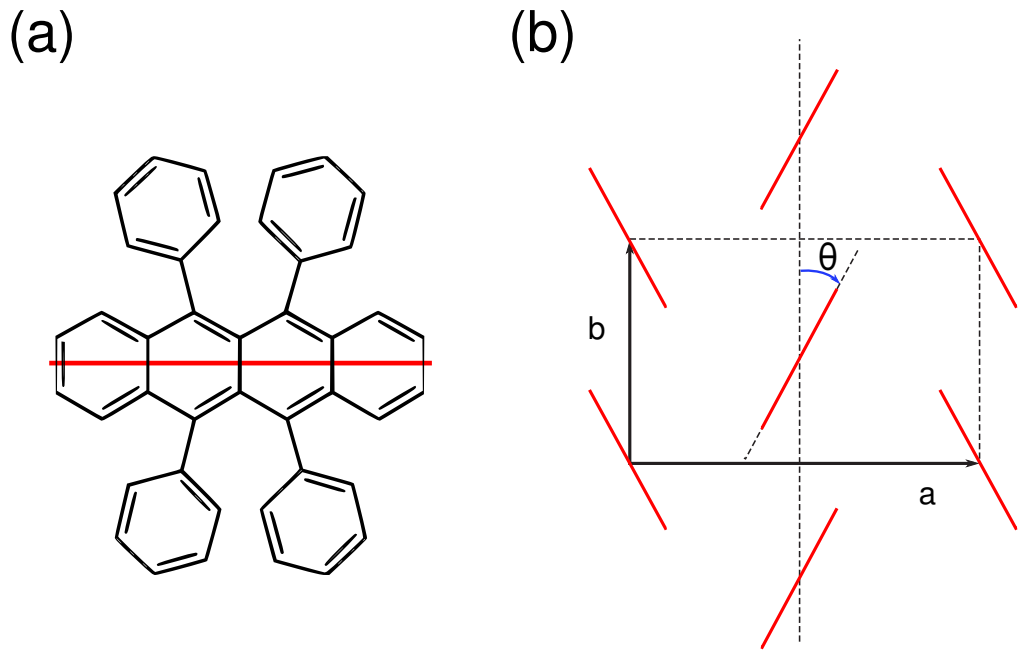}
\caption{\label{figrub} (a) Molecular structure of rubrene. (b) Crystal structure along the (a,b) crystallographic plane.
The red lines denote the long axis of the molecule.}
\end{figure}

Since we are mostly interested in $dc$ conductivity and low frequency
spectral properties, we need to determine an effective Hamiltonian
for the electron degrees of freedom valid for low energy and particle density. We start from the orthorhombic
lattice of rubrene with two molecules per unit cell and  $a$, $b$,
$c$ lattice parameter lengths along the three crystallographic
vectors of the conventional cell \cite{corop}.
We follow the ARPES experiments by Ding {et al.} \cite{Ostrogota} in order to extract the transfer integrals of the highest occupied molecular
orbital (HOMO) bands. The dispersion law of the lowest HOMO is accurately fitted with nearest neighbors tight binding parameters
for small values of momentum providing the following estimates: $t_{a}=118.6meV$ and $t_{b}=68.6meV$ \cite{fernando}. For $t_{c}$ there is
no experimental measure but theoretical estimates seem to agree that
it has to be small compared with other directions owing to the large interplanar separation of rubrene. In the following,
we assume $t_{c}$ much smaller than $t_{a}$ and $t_{b}$ (in the following we assume $t_c=0.18$ $t_{a}$ ). The total volume is $V=L_{a}*L_{b}*L_{c}$, and $L_i$ size along the axis $i=a,b,c$.
We consider two crystalline layers along c because in OFET the effective channel of
conduction covers only few planes \cite{Shehu}.

In Eq.(\ref{h}), the whole lattice dynamics is ascribed to an effective phononic mode whose frequency
$\omega_{0}=\sqrt{K/M}$ is assumed of the order of $5-6meV$ \cite{Troisi Rubrene}. This assumption provides
the low adiabatic ratio $\hbar\omega_0/t_{a}\simeq0.05$ (adiabatic regime).

The electron-phonon interacton $H_{el-ph}$ in Eq.(\ref{h}) is
\begin{equation}
H_{el-ph}=\sum_{\vec{R}_i, \vec{\delta}} \alpha_{|\vec{\delta}|} \left( x_{\vec{R}_i}- x_{\vec{R}_i+\vec{\delta}} \right)c_{\vec{R}_i+\vec{\delta}}^{\dagger}c_{\vec{R}_i}, \quad\vec{\delta}=\vec{a},\vec{b},\vec{c},\label{eq:Tight binding SSH}
\end{equation}
where $\alpha_{\vec{\delta}}$ is electron-phonon parameter controlling the effect
of the ion displacements in the direction $\vec{\delta}$ on the transfer integral. Once fixed
$\alpha_{a}$, we impose $\alpha_{b}/\alpha_{a}=t_{b}/t_{a}$ and,
in the same way, $\alpha_{c}/\alpha_{a}=t_{c}/t_{a}$. The dimensionless quantity
$\lambda$
\begin{equation}
\lambda=\frac{\alpha_{a}^{2}}{4kt_{a}}\label{eq:coupling}
\end{equation}
is the relevant parameter to quantify the electron-phonon coupling strength.
By comparing ab-initio calculations  and average properties of  a simple 1D model, Troisi et al. provided an estimate of   $\lambda \simeq 0.087$\cite{Troisi Rubrene, Bologna ab-initio,Bologna Raman}.
However, in our 3D model the 1D estimate is not correct. Indeed, the hopping $t_b$ is about half $t_a$ and the average kinetic energy is, then, larger with respect to the one-dimensional case. In order to reproduce the same effective one dimensional $\lambda$, we have chosen the larger value $\lambda=0.12$.

In the following part of the section, we use units such that lattice parameter $a=1$, Planck constant $\hbar=1$,
Boltzmann constant $k_B=1$, and electron charge $e=1$.

\subsection{Calculation method}

For the organic semiconductors studied in this paper, the adiabatic ratio is very low implying that the adiabatic limit is appropriate for studying low-frequency intermolecular modes. Consequently, it is possible to adopt a   semiclassical approach: the electron dynamics is fully quantum,
while the ion dynamics is assumed classical. This assumption will limit the temperature range where the results can be considered valid, indeed all the results are valid for temperatures $T \geq \omega_0 \simeq 100 K $. Finally, we will focus on the weak to intermediate electron-phonon regime that seems to be appropriate for high mobility organic semiconductors \cite{corop}.

Within the adiabatic regime, the calculation is equivalent to the classical problem of quantum particles in the presence of an external disordered potential given by the ion displacements $\{ x_{\vec{R}_i} \}$ and controlled by electron-phonon coupling. Each configuration of ion displacements is generated according to the probability function of the  $P \left( \{ x_{\vec{R}_i} \} \right)$, that has to be self-consistently calculated as a function of electron-phonon coupling, temperature and particle density $n=N_p/V$, with $N_p$ number of particles. The adiabatic approach has been also used for the study of molecular junctions and carbon nanotubes at thermodynamical equilibrium and in non-equilibrium conditions \cite{alberto1,alberto2,alberto3,mepumping1,mepumping2}.

In most OFET, the induced doping is not high, therefore, in the following, we will focus on the regime of low doping (up to $n=0.01$). For this regime of parameters, the probability function of the lattice displacements $P \left( \{ x_{\vec{R}_i} \} \right)$ shows very tiny deviations from the distribution of free oscillators \cite{vittoriocheck}, which is therefore used in the following subsection.
Quantities, like spectral function, density of states, and conductivity are calculated through exact diagonalization of the resulting electronic problem at fixed displacements $\{ x_{\vec{R}_i} \}$ and through Monte-Carlo approach for the integration over the distribution $P \left( \{ x_{\vec{R}_i} \} \right)$.
In the case of conductivity, for each configuration of the lattice displacements, we calculate the exact Kubo formula \cite{mahan}
\begin{equation}
Re\left[\sigma_{\rho,\rho}\left(\omega\right)\left(\left\{ u_{\mathbf{R}_{i}}\right\} \right)\right]=  \frac{\pi\left(1-e^{-\beta\omega}\right)}{V\omega}\sum_{r\neq s}p_{r}\left(1-p_{s}\right)\nonumber \\
 \left|\left\langle r\left|J_{\rho}\right|s\right\rangle \right|^{2}\delta\left(E_{s}-E_{r}+\omega\right),\label{eq:Kubo formula}
\end{equation}
where $\rho=a,b,c$, $\beta=1/T$, and $p_{r}$ is the Fermi distribution
\begin{equation}
p_{r}=\frac{1}{1+\exp\left(\beta\left(E_{r}-\mu_{p}\right)\right)}\label{eq:Fermi factor}
\end{equation}
corresponding to the exact eigenvalue $E_{r}$ at any chemical potential
$\mu_{p}$. Finally $\left\langle r\left|J_{\rho}\right|s\right\rangle $
is the matrix element of the current operator $J_{\rho}$ along the
direction $\hat{e}_{\rho}$, defined as
\begin{equation}
J_{\rho}=i\sum_{\vec{R}_{i},\vec{\delta}}\bar{t}_{\vec{\delta}}\left(\mathbf{R}_{i}\right)
\left(\vec{\delta}\cdot\hat{e}_{\rho}\right)c_{\vec{R}_{i}}^{\dagger}c_{\vec{R}_{i}+\vec{\delta}},
\label{current}
\end{equation}
with
\begin{equation}
\bar{t}_{\vec{\delta}}\left(\mathbf{R}_{i}\right)=t_{\vec{\delta}}-\alpha_{\vec{\delta}}
\left(u_{\mathbf{R}_{i}}-u_{\mathbf{R}_{i}+\vec{\delta}}\right), \quad\vec{\delta}=\vec{a},\vec{b},\vec{c}.
\end{equation}
We notice that, in contrast with spectral properties, the temperature enters the calculation not only through
the displacement distribution, but also directly for each configuration
through the Fermi distributions $p_{r}$. We point out that the current-current correlator is not evaluated at the lowest order as convolution of two single-particle Green functions, but the linear response conductivity is exactly calculated within the Kubo formulation  presented in Eq. (\ref{current}) \cite{mahan}. Therefore, the numerical calculation of the conductivity is able to include the vertex corrections (terms in the correlator beyond the convolution of two Green functions) discarded by previous approaches \cite{mahan}.

Finally, the mobility $\mu$ is calculated as the ratio between zero-frequency conductivity and carrier density:
\begin{equation}
\mu_{\rho}=\lim_{\omega\rightarrow0^{+}}\frac{Re\left[\sigma_{\rho,\rho}\left(\omega\right)\right]}{n}.\label{eq:Mobility}
\end{equation}

Summarizing, the  numerical method provides approximation-free results in the adiabatic regime. The only limitation is due to the computational time being controlled by matrix diagonalizations.

\subsection{Results about spectral and transport properties}

\begin{figure}[htb]
\centering
\includegraphics[width=0.75\textwidth]{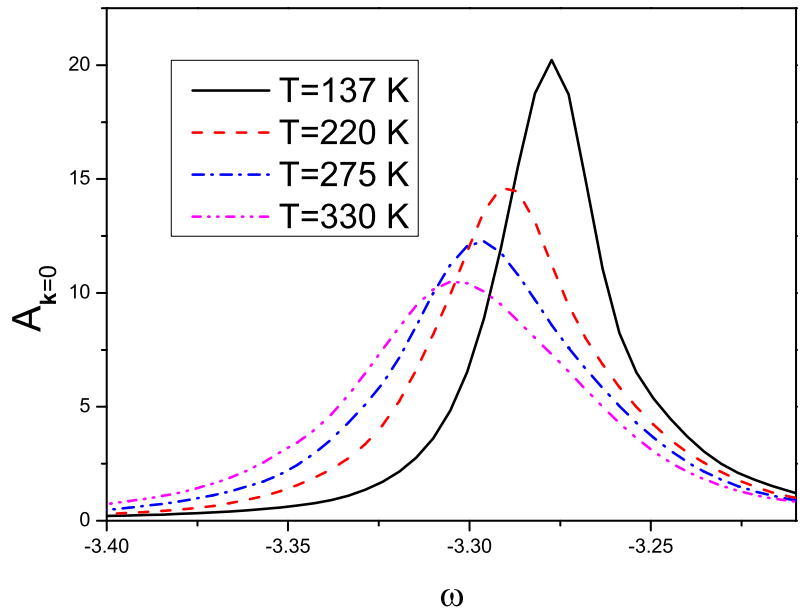}
\caption{\label{fig1} The spectral function (in units of $1/t_a$) for rubrene model at momentum $\mathbf{k}=0$ as
a function of the frequency (in units of $t_a/\hbar$) at $\lambda=0.12$ and $n=0.002$ for different temperatures T.
We assume the hopping parameter $t_c=0.18 t_a$, with $t_{a}=118.6meV$.}
\end{figure}

Spectral and transport properties, studied within the two models based on Eq. (\ref{h}), will be discussed in this subsection.
The study of spectral properties is important to individuate the states that mainly contribute to the conduction process. The spectral properties within the anisotropic Holstein model bear strong resemblance with the rubrene model, so that we will discuss only this last model.

In Fig. \ref{fig1}, we report the spectral function at momentum $\mathbf{k}=0$ for the model
parameters of rubrene. We point out that states close to
$\mathbf{k}=0$ are weakly damped. Moreover, with increasing temperature, the peak position
of the spectral function is only poorly renormalized in comparison
with the bare one, in agreement with results of 1D SSH model \cite{vittoriocheck}. Therefore, these states keep the itinerant character of the bare ones, and they will be involved into the diffusive conduction process (see discussion in the next paragraph). We notice that the spectral function at $\mathbf{k}=0$ is extremely small at the chemical potential
$\mu_{p}$ for all the temperatures. For example, at $T=275$ K, the spectral weight is concentrated in
an energy region higher than that in which the chemical potential
is located ($\mu_{p}=-3.74 t_a$ for  $n=0.002$). Actually, the spectral
weight is mainly in the region between $\mu_{p}+2T$ and $\mu_{p}+3T$.

\begin{figure}[htb]
\centering
\includegraphics[width=0.75\textwidth]{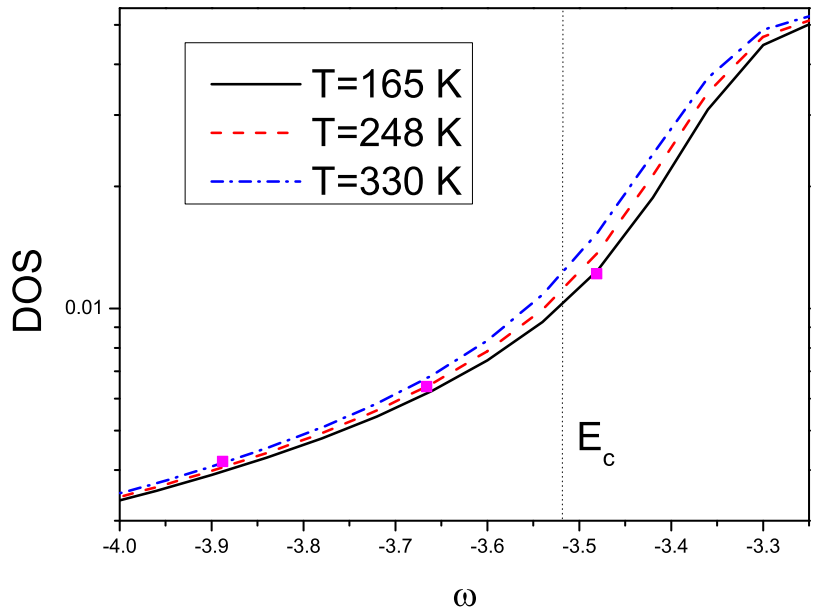}\caption{\label{fig2} The DOS (in units of $1/t_a$) for rubrene model as a function of the frequency (in units of $t_a/\hbar$) at $\lambda=0.12$ and $n=0.002$ for different temperatures T. The squares indicate the chemical potential $\mu_{p}$ (in units of $t_a$) at fixed temperature. $E_{c}$ (dot line, in units of $t_a$) is the free electron band edge close to the mobility edge. We assume the hopping parameter $t_c=0.18 t_a$, with $t_{a}=118.6meV$. }
\end{figure}

We have checked that the spectral functions with low momentum are more
peaked, while, with increasing $\mathbf{k}$, they tend to broaden. The density of states (DOS) can be calculated as the sum of the spectral
functions $A_{\mathbf{k}}$ over all the momenta $\mathbf{k}$. The tail in the DOS is due to the marked width
of the high momentum spectral functions. In Fig. \ref{fig2},
the DOS is shown for different temperatures at
$\lambda=0.12$ and $n=0.002$. As shown in logarithmic scale, the
DOS has a tail with a low energy exponential behavior. This region
corresponds to localized states \cite{economou}. We have, indeed, checked
analyzing the wave functions extracted from exact diagonalizations
that, actually, states with energies deep in the tail are strongly
localized (one or two lattice parameters along the different directions
as localization length). On the other hand, close to the shoulder ($\omega \simeq -3.4 t_a$, see
Fig. \ref{fig2}), the itinerant nature of states is clearly obtained.
This analysis allows to give an estimation for the mobility edge energy (the energy that divides localized and itinerant states)  which can be
located very close to the band edge $E_{c}$ for free electrons (in our case, $E_{c}=-3.52 t_a$).

The number of localized states available in the tail increases with temperature. It is
important to analyze the role played by the chemical potential $\mu_{p}$
with varying the temperature. Actually, $\mu_{p}$ enters the energy
tail and will penetrate into it with increasing temperature. At fixed
particle density $n=0.002$, for $T=165$ K, one has $\mu_{p}=-3.49 t_a$,
while, for $T=330$ K, $\mu_{p}=-3.88 t_a$ (see squares of Fig. \ref{fig2} for the values of the chemical potential).
One important point is that the quantity $E_{c}$ and the close mobility
edge are significantly larger than $\mu_{p}$. Therefore, in the regime of low density
relevant for OFET, the itinerant states are not at $\mu_p$, but at higher
energies. We point out that those are the states relevant for the conduction
process. Therefore, the analysis of the properties of a high dimensional
model points out that both localized and itinerant states are present
in the system. This is a clear advantage of our work over previous
studies in low dimensionality \cite{Troisi Orlandi,Troisi2D,Ciuchi Fratini}
in which all states are localized: more localized at very low energy
and less localized close to the free electron edge. Summarizing,
in our system, with increasing temperature, all the states up to $\mu_{p}$ become localized and the itinerant states
become statistically less effective due to the behavior of the chemical
potential. Eventually, the effect of penetration of $\mu_{p}$ in
the tail, due to the Fermi statistics, will overcome the effects of available itinerant states around $E_c$. We have checked that the penetration of the
chemical potential towards the energy region of the tail is enhanced
with increasing the electron-phonon coupling.

\begin{figure}[htb]
\centering
\includegraphics[width=0.56\textwidth,angle=0]{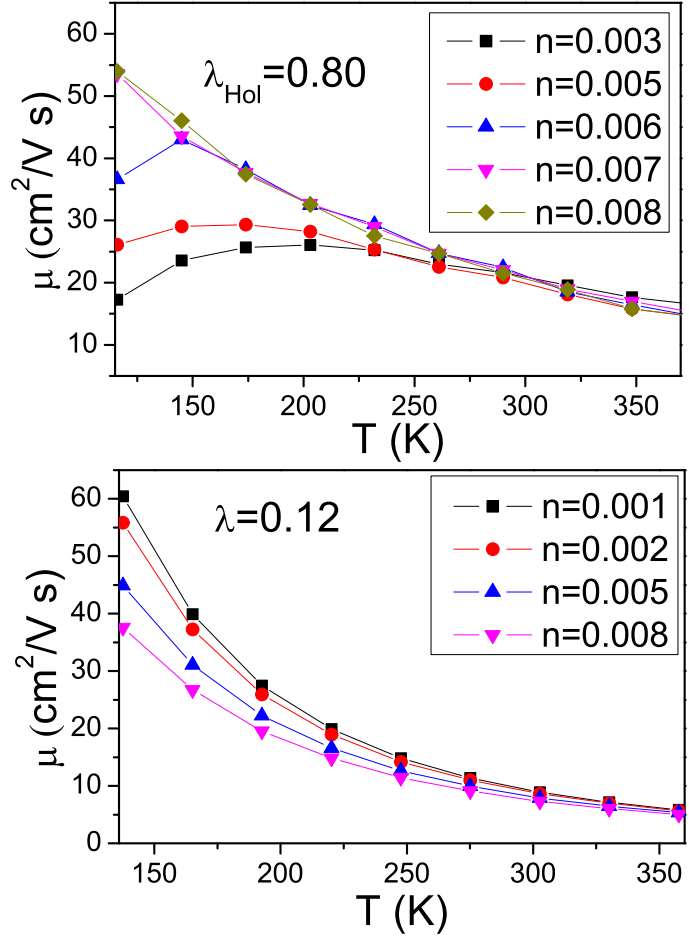}
\caption{\label{fig3} Upper Panel: Mobility along the z direction within the anisotropic Holstein model as a function of temperature T for different particle densities at $\lambda_{Hol}=0.8$.
Lower panel: Mobility along the $a$ direction within the rubrene model as a function of temperature T for different particle densities at $\lambda=0.12$. We assume the hopping parameter $t_c=0.18 t_a$, $t_{a}=118.6meV$.}
\end{figure}

We devote the last part of the subsection to analyze the transport properties in both the anisotropic Holstein model and the rubrene model.
As shown in the upper panel of Fig. \ref{fig3},
we first discuss the mobility of the anisotropic Holstein model along $z$ direction as a function of temperature with changing the particle density. Within this model, the carrier density strongly affects the behavior of mobility showing a cross-over from metallic to insulating behavior at low temperatures. For densities around one per cent, the chemical potential at low temperature is above the mobility edge and, then, the mobility is metallic-like. On the other hand the situation is different for densities below one per cent. At those densities, the chemical potential is below the mobility edge, so that mobility shows an insulating character. Translated in a polaronic framework, this would mean that the single particle sees a potential well due to electron-phonon coupling, hence, at very low densities, a polaron-like activated mechanism could set in. Actually, the mobility results are consistent with those obtained within the picture of polaron formation \cite{hannewald}.

The strong charge density dependence on the mobility is, instead, lost at higher temperatures where the chemical potential is always in the tail regions for all the charge densities studied, but more states at higher energies (of itinerant nature) get involved providing the main contribution to mobility. As a consequence of the different behavior with charge densities at low and high temperatures, the mobility changes its character at intermediate temperatures for low charge carrier densities.

At high temperatures, the mobilities look very similar for all the densities considered in the upper panel of Fig. \ref{fig3}. At room temperature, the mobility is about $20 cm^2 /(V \cdot s)$, a value that recovers the right order of magnitude of experimental data in oligoacenes \cite{morpurgo}. However, the mobility decreases with the temperature as $1/T$, not in agreement with measurements in oligoacenes, such as rubrene. Moreover, in this model, polaronic localization seems to take place in the low temperature range for enough low charge densities even if experimental data do not seem to support this scenario. Therefore, a more accurate model for the electron-phonon coupling is needed. As already discussed in subsection $2.2$, the SSH-like coupling with intermolecular vibrational modes is what we need. In the following part of this subsection, we will analyze the transport properties of that model.

In the lower panel of Fig. \ref{fig3}, the mobility along the
$a$ direction is reported as a function of the temperature at fixed
coupling $\lambda=0.12$ and different concentrations $n$.
The plot shows that the absolute magnitude of the mobility substantially
agrees with the experimental estimates being $\mu\simeq10$ $cm^{2}/(V s)$
at room temperature. Furthermore, the mobility exhibits a band-like power-law $T^{-\gamma}$ behaviour for all the concentrations.
The exponent $\gamma$ is evaluated from fits of the mobility providing values
in the range $2-2.4$, where the highest value is related to the
lowest concentration. This trend is in agreement with experimental
measures that for rubrene establish $\gamma\simeq2$ for temperatures
$T>170-180K$ \cite{morpurgo}. A feature, in contrast with the Holstein model, is that the
mobility increases with decreasing the concentration of carriers. This trend,
already found in 1D SSH model \cite{vittoriocheck}, points out that the there is no room for a polaronic (bond) localization \cite{capone}
within the regime of rubrene parameters explored in this review.

Another important property is the anisotropy of the transport properties along different in-plane directions \cite{hanne1}. In our model for rubrene, the anisotropy of the mobility along different cristallographic directions is essentially the same as the anisotropy of the effective mass. From the estimates of in-plane hoppings $t_a$ and  $t_b$, the anisotropy of the mobility is evaluated to be of the order of $(t_b/t_a)^2= 0.335$. From experiments \cite{morpurgo}, the anisotropy ratio is about $0.375$ at room temperature, therefore in agreement with our estimate. However, the experimental data shows that this ratio increases with decreasing temperature. In a recent paper \cite{ishii}, where the electron-phonon coupling is slightly more complex than the SSH-like interaction considered in our model, the anisotropy of the mobility is less marked than that of the mass. Within this paper, the anisotropy ratio between the two in-plane mobilities is calculated to be of the order of 0.44 at room temperature (higher than the experimental value), and 0.5 at T=150 K, therefore, it increases with decreasing temperature in agreement with experimental trend. Even if the electron-phonon models are slightly different, the conclusions of our work and these recent calculations are qualitatively consistent. Finally, it has been suggested \cite{hanne6} that the change in mobility anisotropy upon temperature variation can be explained by a change in transport characteristics (from band transport to hopping).

Starting from the mobility, we can determine the scattering time $\tau_{tr}$ from
the relation $\mu=e\tau_{tr}/m$. Since the mass is weakly renormalized
from the electron-phonon interaction, one can assume $m$ as the bare mass at
$\mathbf{k}=0$. We point out that $\tau_{tr}$ is on the scale of the fs, so that it is one order of magnitude
lower than the damping time of the states important for the spectral
properties (on the scale of ten fs). Therefore, the transport processes
amplify the effects of the electron-phonon interaction and the vertex corrections
introduced within our approach are fundamental to take into account
this effect.

From the scattering time, one can deduce the mean free path as $l_{tr}=v_{av}\tau_{tr}$,
where $v_{av}$ is the average velocity of the charge carriers.
The quantity $l_{tr}$ is always on the scale of a few lattice parameters.
The most important feature is its temperature
behavior. As a consequence of the electron-phonon effects, close
to room temperature, it becomes of the order of half lattice parameter
$a$. This means that the Ioffe-Regel limit is reached \cite{gunnarsson}.
The decrease of the mobility in the Ioffe-Regel limit is not due to
a mass renormalization (dynamic and/or static) but it is due to a reduction
of the available itinerant states (the only ones able to transport
current) with the temperature. We remark that this result is due to
the fundamental role played by vertex corrections (introduced in the previous subsection about the computational methods)
in the calculation of the mobility.


\section{Effects of combined low frequency inter-molecular and high frequency intra-molecular vibrational modes}

For rubrene and pentacene, the carrier mobility is dominated by inter-molecular phonons since the interaction with intra-molecular modes is almost negligible \cite{Troisi Rubrene} and, then, the model discussed in the previous section is considered adequate. On the other hand, in other oligoacenes with a smaller number of benzene rings, the coupling with local modes cannot be neglected \cite{corop}.

Indeed, the decrease of the number of benzene rings affects the reorganization energy, which can be related to the binding energy of the polaron, that is the quasi-particle formed by the electron (or hole) and the surrounding phonon cloud.
Actually, going from pentacene to naphthalene, the reorganization energy increases nearly twice suggesting a much stronger coupling with local modes \cite{corop}. Therefore, for systems with reduced number of benzene rings like naphthalene, we expect a larger interplay between intra- and inter-molecular modes (see Fig. \ref{naphta}) within the intermediate electron-phonon coupling regime for both modes. The next step of the review is to combine the effects of high frequency local vibrational (antiadiabatic) modes with non local low frequency (adiabatic) ones \cite{meholpssh}. For computational ease, we will restrict our analysis to the 1D case assuming the inter-molecular phonons classical, but considering the intra-molecular modes fully quantum.

\subsection{Model hamiltonian}

We consider a one-dimensional model with coupling to intra- and inter-molecular modes \cite{meholpssh,piegari} similar to one recently introduced, where the treatment only concerns the study of spectral properties \cite{ciuchi}.
The coupling to intra-molecular modes is Holstein-like, that to inter-molecular modes is SSH-like (see Fig \ref{naphta} for a sketch of the two couplings).
It can be summarized in the following hamiltonian:
\begin{equation}
H= H_{el}^{(0)}+H_{Intra}^{(0)}+H_{Inter}^{(0)}+H_{el-Intra}+H_{el-Inter}.
\label{hgen}
\end{equation}

In Eq. (\ref{hgen}), the free electronic part $H_{el}^{(0)}$ is
\begin{equation}
H_{el}^{(0)}=-t \sum_{i}  \left( c_{i}^{\dagger}c_{i+1}+ c_{i+1}^{\dagger}c_{i}   \right),
\label{helgen}
\end{equation}
where $t$ is the bare electron hopping between the nearest neighbors on the chain, $c_{i}^{\dagger}$ and
$c_{i}$ are the charge carrier creation and annihilation operators, respectively, relative to the site $i$ of a chain with lattice parameter $a$.
For the transfer hopping the {\it ab-initio} estimate is: $t \simeq 50-100 meV$ \cite{corop}. We consider a single-band one-dimensional electronic structure since it represents the simplest effective model in anisotropic organic semiconductors to analyze the low energy features responsible for the mobility properties.

In Eq. (\ref{hgen}), $H_{\alpha}^{(0)}$, with $\alpha= Intra, Inter$, is the Hamiltonian of the free optical molecular modes
\begin{equation}
H_{\alpha}^{(0)}= \sum_{i} \frac{{p}^2_{\alpha,i}}{2 m_{\alpha}}+ \sum_{i} \frac{ k_{\alpha} x_{\alpha,i}^{2}}{2},
\label{hintragen}
\end{equation}
where $p_{\alpha,i}$ and ${x}_{\alpha,i}$ are the oscillator momentum and position of the mode $\alpha$, respectively, $m_{\alpha}$ the oscillator mass and $k_{\alpha}$ the elastic constant of the mode $\alpha$. The inter-molecular modes are characterized by small frequencies ($\hbar \omega_{Inter} \simeq 5-10 meV$) in comparison with the transfer hopping \cite{corop,Troisi Orlandi}. On the contrary, the most coupled intra-molecular modes have large frequencies ($\hbar \omega_{Intra} \simeq 130-180 meV$) \cite{corop}.

In Eq. (\ref{hgen}), $H_{el-Intra}$ is the Holstein-like Hamiltonian describing the electron coupling to
intra-molecular modes
\begin{equation}
H_{el-Intra}= \alpha_{Intra} \sum_{i} x_i n_i,
\label{hcoupling}
\end{equation}
with $\alpha_{Intra}$ coupling constant to local modes and $n_i=c_{i}^{\dagger}c_{i}$ local density operator. The dimensionless constant
\begin{equation}
g_{Intra}= \alpha_{Intra} / \sqrt{2 \hbar m_{Intra} \omega_{Intra}^3}
\end{equation}
is used to describe this electron-phonon coupling \cite{holstein1}. In single crystal
organic semiconductors, $g_{Intra}$ is in the weak to intermediate regime (of the order of unity) \cite{corop}.

Finally, in Eq. (\ref{hgen}),  $H_{el-Inter}$ represents the SSH-like term with electron coupling to inter-molecular modes
\begin{equation}
H_{el-Inter}= \alpha_{Inter} \sum_{i} (y_{i+1}-y_i) \left( c_{i}^{\dagger}c_{i+1}+ c_{i+1}^{\dagger}c_{i}   \right),
\label{hcouplinggen1}
\end{equation}
with $\alpha_{Inter}$ coupling constant to non local modes. In the adiabatic regime for non local modes ($\hbar \omega_{Inter} \ll t$), the dimensionless quantity
\begin{equation}
\lambda_{Inter}=\alpha_{Inter}^2/4 k_{Inter} t
\end{equation}
fully provides the strength of the electron coupling to inter-molecular modes. The typical values of $\lambda$ are in the intermediate (of the order of $0.1$) coupling regime \cite{fernando}.

In the following part of this section, we will use units such that lattice parameter $a=1$, Planck constant $\hbar=1$, Boltzmann constant $k_B=1$, and electron charge $e=1$. We will analyze systems in the thermodynamic limit and we will measure energies in units of
$t \simeq 80 meV$.  We fix $\omega_{Intra}=2.0 t$ as model parameter with the highest energy \cite{corop}.

\subsection{Calculation method}

Since a very low carrier density is injected into the organic semiconductor, we will study the case of non interacting particles. The temperature range where intrinsic effects are relevant is $\omega_{Inter} \leq T \ll t < \omega_{Intra} $. Therefore, the dynamics of intermolecular modes can be assumed classical.
On the other hand, it is important to retain the quantum nature of high frequency local vibrational modes.

Actually, the electron motion is strongly influenced by the statistical "off diagonal" disorder, that, in the limit of low carrier density, is described by the probability function $P \left( \{ y_j \}  \right) $ of free classical harmonic oscillators. At a fixed configuration of non local displacements $\{ y_j \}$, Eq. (\ref{hgen}) is equivalent to a Holstein model with displacements $\{ x_i \}$, where the electron hopping depends on the specific nearest neighbor sites  throughout the assigned $\{ y_j \}$.  The resulting inhomogeneous Holstein model can be accurately studied within the modified variational Lang-Firsov approach via a unitary transformation $U \left( \{ y_j \}  \right)$, depending on the non local displacements $\{ y_j \}$ and appropriate in the anti-adiabatic regime ($\omega_{Intra} > t $) \cite{lang,memanganiti}. The electron mass is renormalized by the coupling with local modes (polaronic effect), and the Holstein-coupled oscillators  $\{ x_i \}$ are displaced from their equilibrium position to a distance proportional to the electron-phonon interaction. For each fixed configuration $ \{ y_j \}$, one has to calculate quantities, such as spectral function, density of states, and mobility (calculated as the ratio between conductivity and carrier density), within the Lang-Firsov approach. Then, the effect of non-local adiabatic inter-molecular modes can be taken into account  making the integral over the distribution $P \left( \{ y_i \} \right)$ by means of a Monte-Carlo procedure.

The method exposed above is very accurate in the regime $\omega_{Inter} \ll t$ and $\omega_{Intra} > t $ appropriate to high-mobility organic semiconductors. It properly takes into account the quantum effects of high frequency local vibrational modes. Moreover, the approach is able to include spatial correlations relevant in quasi one-dimensional systems, in particular vertex corrections in the calculation of mobility.

\subsection{Results on spectral and transport properties}

\begin{figure}[htb]
\centering
\includegraphics[width=0.75\textwidth,angle=0]{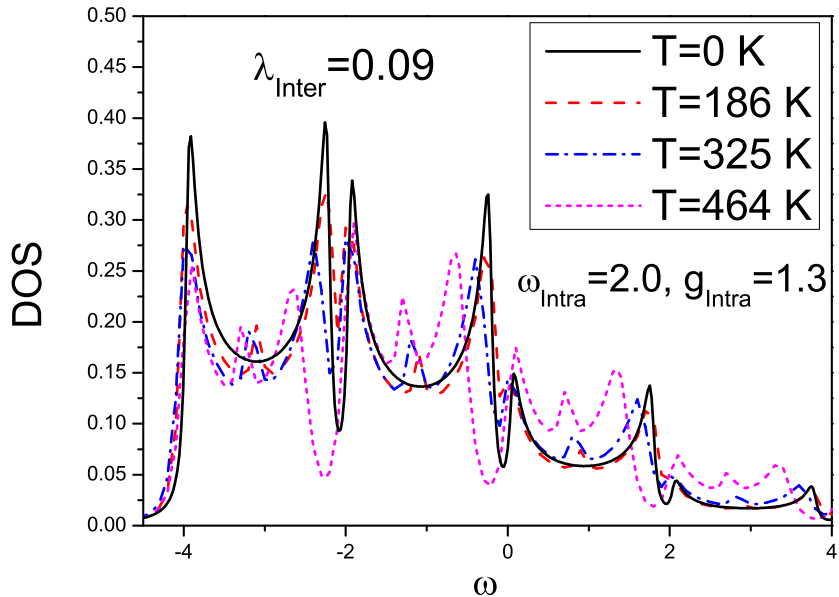}
\caption{The DOS (in units of $1/t$) as a function of the frequency (in units of $t/\hbar$)  for different temperatures T at $\lambda_{Inter}= 0.09$ and $g_{Intra}=1.30$. We consider $\omega_{Intra}=2t$ (in units of $t$, with $t= 80 meV$).}
\label{fig4}
\end{figure}

In Fig. \ref{fig4}, we report the density of states (DOS) for $g_{Intra}=1.3$ and $\lambda_{Inter}=0.09$ at different temperatures.
For all the temperatures, there is a strong renormalization of the bare band whose width becomes twice smaller (from $4t$ to roughly $2t$) and move to lower energies (from $-2t$ to roughly $-4t$). Furthermore high energy satellite bands appears at multiples of the vibrational frequency $\omega_{Intra}=2t$ \cite{mahan} providing a DOS extending from $-4t$ to $4t$. These effects can be easily ascribed to the local modes since they survives also at $T=0$, where the effect of non local modes is weak.
 The intrisic reduction of the bare band due to local modes provides a simple and direct explanation of the difference in the bandwidth evidenced in the series of oligoacenes from naphthalene (effective band of the order of $40 meV$) to pentacene (effective band of the order of $80 meV$). Indeed, it can be ascribed to the decrease of the reorganization energy with increasing the benzene rings of the single molecules that in turn reduces the renormalization effects \cite{corop}. Within the polaron theory \cite{mahan}, the narrowing of the main band is related to the spectral weight Z of the quasi-particle, which is estimated to be about $0.5$ from the calculations. Therefore, our estimate of Z compares favorably with recent ab-initio results for which Z relative to  the electron channel is of the order of 0.7 \cite{vukmi}.

At finite temperature, the shape of the spectra is changed due to the non local coupling. Actually, any band shows a new small maximum due to the coupling to inter-molecular modes. It is well known that, in the polaron theory, the band narrowing increases strongly with temperature. In order to be more quantitative, we notice that, for $\lambda_{Inter}=0$, the principal band at $T=325$ K is reduced of about $42 \%$ of the band at $T=0$. On the other hand, for  $\lambda_{Inter}=0.09$, the principal band at $T=325$ K is reduced of only $7 \%$ of the band at $T=0$. Unexpectedly, in our model, the band narrowing is strongly reduced due to the non local coupling.  The narrowing of the principal band results from a subtle equilibrium between the two opposite tendencies. Actually, the coupling to non local modes has the main effect to induce scattering into the single bands of the density of states, preventing the narrowing induced by the coupling to local modes.
The interplay between local and non local modes is able to produce a modest narrowing as function of the temperature even if the coupling to local modes is not
weak. Our prediction is that this effect should be present not only in pentacene \cite{arpes2},  but also in naphthalene and anthracene.

\begin{figure}[htb]
\centering
\includegraphics[width=0.75\textwidth,angle=0]{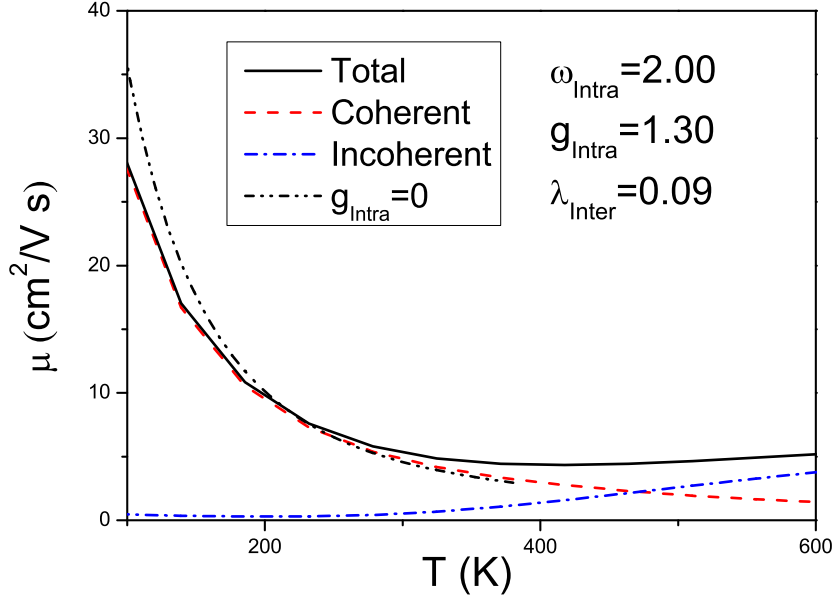}
\caption{Mobility and its different contributions as a function of the temperature T at $\lambda_{Inter}= 0.09$ and $g_{Intra}=1.3$.
We consider $\omega_{Intra}=2$ (in units of $t$, with $t= 80 meV$).}
\label{fig5}
\end{figure}

Next, we analyze the mobility in the intermediate regime for both intra- and inter-molecular modes (see Fig. \ref{fig5}). The mobility can be divided into two contributions: the coherent one, where the scattering of the renormalized electron (the only effect due to local electron-phonon coupling is here the reduction of the bandwidth)  with non local modes is included, and the incoherent one, where, in addition to non local modes, scattering with multiple real local phonons is considered.

The coherent term of mobility, relevant at low temperatures, bears a strong resemblance with the mobility of the system at $g_{Intra}=0$, even if, as expected, it is smaller.  The local coupling is able to affect but to not destroy the low temperature behavior dominated by the non local coupling. Actually, for $g_{Intra}=0$, the mobility scales as $1/T^{1.89}$, while, with increasing $g_{Intra}$, the power-law becomes slightly less pronounced. In the case $g_{Intra}=1.3$, the mobility goes as $1/T^{1.60}$, still compatible with experiments in naphthalene and it has the correct order of magnitude \cite{warta}.

The incoherent term of mobility starts at a temperature of about $T=230$ K and becomes predominant only at temperatures much higher than room temperature. The role of the local coupling here is to promote an activated behavior in the incoherent regime which is effective only at high temperature. Consequently, the local coupling provides a negligible contribution to the mobility up to room temperature.  Actually, the combined effect between intra- and inter-molecular modes is able to provide an activation energy $\Delta$ of only about $20 meV$, therefore less than one half of that for $\lambda_{Inter}=0$ and close to that extracted by experimental data in naphthalene (about $15$ meV for mobility along c-axis) \cite{warta}. We stress that the small activation energy is found even if the reorganization energy related to intra-molecular modes derived from ab-initio calculations \cite{corop} is not small. Actually, the polaronic binding energy is given by $g_{Intra}^2 \omega_{Intra}$, which is of the order of $1.7*2*t \simeq 270$ meV. Therefore, the non-local SSH interaction is able to strongly quench the tendency towards localization of the local Holstein coupling. As a result, the activation energy of the transport properties is estimated to be much smaller than the value of the polaronic local energy if other electron-phonon non local interactions are playing a relevant role.

Summarizing, the proposed model is able to capture many features of the mobility in oligoacenes.

\section{Effects of gates made of polarizable dielectrics and disorder}

\begin{figure}[htb]
\centering
\includegraphics[width=0.60\textwidth,angle=0]{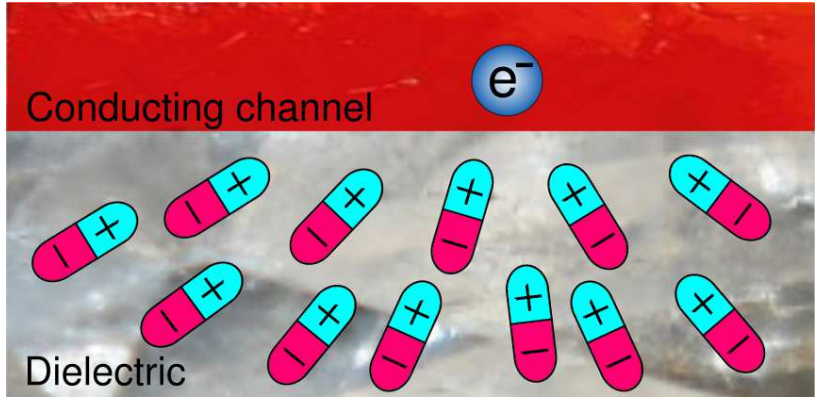}
\caption{Sketch of the effects induced by the charge carrier in the conducting channel on the gate close to the interface. The electron in  the organic semiconductor induces a polarization within the dielectric that, in turn, affects the electron dynamics.}
\label{figdie}
\end{figure}

In the last part of this review, we investigate the effect of a polarizable gate on the transport properties of organic semiconductors \cite{substrato,mebipo} (see Fig. \ref{figdie} for a sketch about the coupling between charge carrier and polarization in the dielectric). This analysis is important to interpret experimental data in rubrene OFET grown on polarizable dielectric gates, such as the $Ta_2 O_5$ oxide \cite{nature}.

\subsection{Model Hamiltonian}

We study a one-dimensional Hamiltonian model with coupling to bulk and interface vibrational modes \cite{substrato}. This model is similar to that of the previous section. Actually, the free electronic hamiltonian is the same, and the bulk modes of this model correspond to the inter-molecular modes.

The model is described by the following Hamiltonian
\begin{equation}
H= H_{el}+H_{Bulk}^{(0)}+H_{el-Bulk}+H_{Int}^{(0)}+H_{el-Int}.
\label{hgen1}
\end{equation}

In Eq. (\ref{hgen1}), the electronic part $H_{el}$ is given by Eq. (\ref{helgen}) of the previous section, with $t$ bare electron hopping (estimated to be among $80$ meV  and $120$ meV) between the nearest neighbors sites.

In Eq. (\ref{hgen1}), $H_{Bulk}^{(0)}$ corresponds to Eq. (\ref{hintragen}) for free intermolecular modes with elastic constant $k$, mass $m$, and $\hbar \omega_{Bulk} \simeq 5-10$ meV much smaller than transfer hopping t \cite{corop,Troisi Orlandi}.

In Eq. (\ref{hgen1}),  $H_{el-Bulk}$ represents the term similar to the SSH \cite{SSH} interaction for the coupling to intermolecular modes given in Eq. (\ref{hcouplinggen1}). As in the previous section, one can define $\lambda_{Bulk}$ whose typical values are in the intermediate coupling regime (in this section, we take the value $\lambda_{Bulk}=0.1$ suitable for rubrene) \cite{vittoriocheck}.

In Eq. (\ref{hgen1}), $H_{Int}^{(0)}$ is the Hamiltonian of free interface phonons
\begin{equation}
H_{Int}^{(0)}= \hbar \omega_{Int} \sum_{q} a_q^{\dagger} a_q,
\label{hintra}
\end{equation}
where $\omega_{Int}$ is the frequency of optical modes, $a_{q}^{\dagger}$ and $a_{q}$ are creation and annihilation operators, respectively, relative to phonons with momentum $q$.

In Eq. (\ref{hgen1}), $H_{el-Int}$ is the Hamiltonian describing the electron coupling to interface vibrational modes
\begin{equation}
H_{el-Int}= \sum_{i,q} M_q n_i e^{i q R_i} \left( a_q + a_{-q}^{\dagger}   \right),
\label{hcouplingel}
\end{equation}
where $ n_i$ is the density operator, $M_q$ is the interaction electron-phonon term
\begin{equation}
M_q= \frac{g \hbar \omega_{Int} }{\sqrt{L}} \sum_{i} e^{i q R_i} \frac{R_0^2}{R_0^2+R_i^2},
\label{hcouplingel1}
\end{equation}
with $g$ dimensionless coupling constant, $L$ number of lattice sites, $R_i$ position of the site $i$, and $R_0$ cut-off length of the order of the lattice spacing $a$. This electron-phonon coupling describes the long-range interaction induced on the electron at the interface with the dielectric gate.
In order to quantify this coupling, we use the dimensionless quantity
\begin{equation}
\lambda_{Int}=\sum_q \frac{M_q^2}{2 \hbar \omega_{Int} t}.
\end{equation}
In this work, we take $R_0=0.5 a$ and $\hbar \omega_{Int}=0.5 t$ \cite{bussac}.

In the following part of this section, we will use units such that $a=1$, $\hbar=1$, $e=1$, and Boltzmann constant $k_B=1$.
We will analyze systems in the thermodynamic limit measuring energies in units of
$t \simeq 100$ meV. The calculation method is similar to that in the previous section since the role of intra-molecular modes is here played by interface modes.

\subsection{Results about transport properties}

\begin{figure}[htb]
\centering
\includegraphics[height=0.50\textwidth,angle=0]{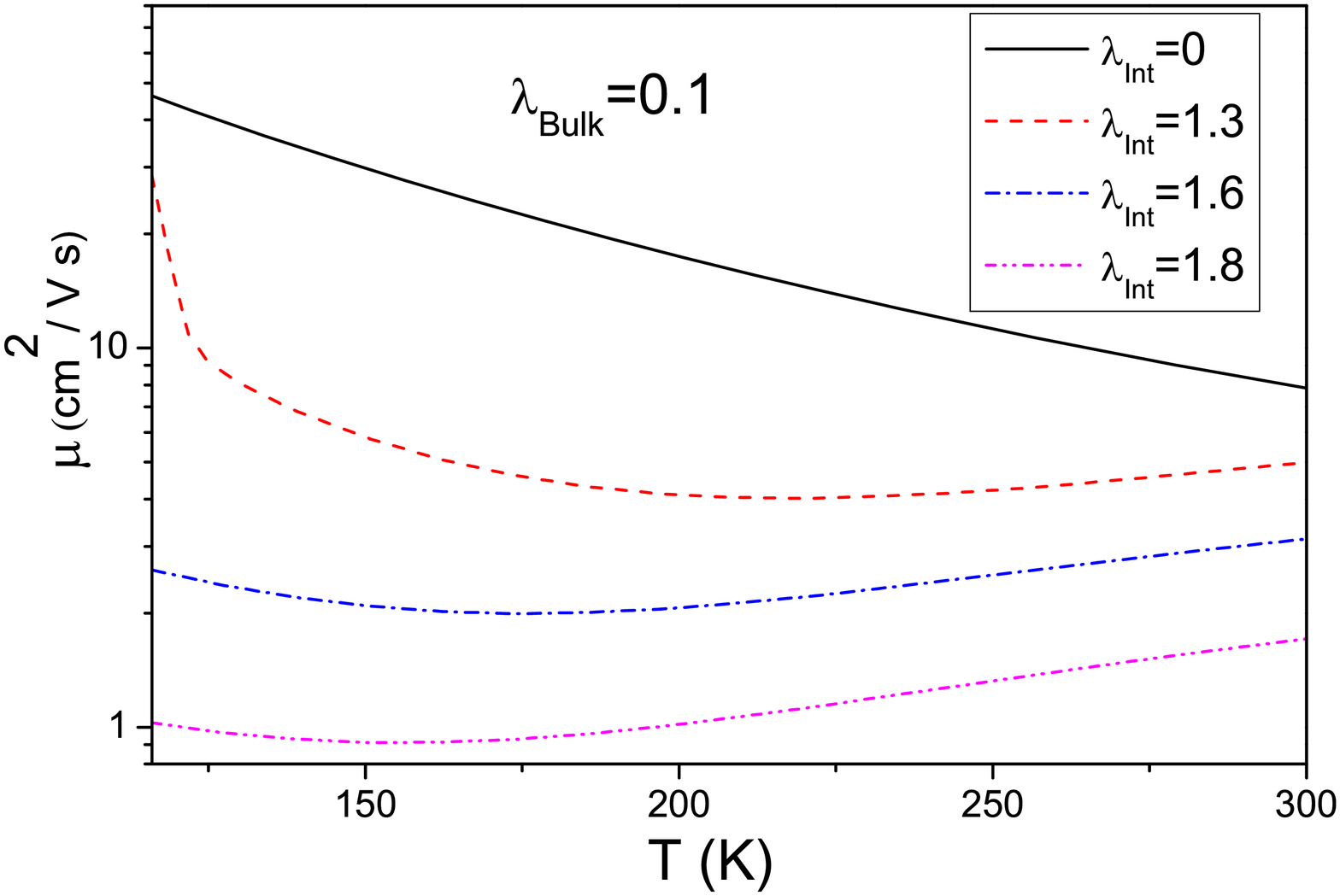}
\caption{Mobility $\mu$ as a function of the temperature T at $\lambda_{Bulk}=0.1$ for different values of $\lambda_{Int}$. }
\label{fig6}
\end{figure}

In Fig. \ref{fig6}, we report the mobility as a function of the temperature for different values of $\lambda_{Int}$ at bulk coupling $\lambda_{Bulk}=0.1$ (appropriate to rubrene). The quantity $\mu$ shows a coherent band-like behavior at low temperatures, but, with increasing $T$, it goes towards the activated behavior where the bulk coupling is not effective. Actually, the mobility interpolates between the behaviors with only bulk and interface phonons. At low temperature, the diffusive contribution is ascribed to the modulation of the electron kinetic energy due to the bulk modes with SSH interactions. This coherent contribution is weakened with increasing $\lambda_{Int}$, but it does not disappear. This element is in contrast with experimental data which show a more or less marked insulating behavior from $150$ K to $300$ K \cite{nature}. Therefore, the theoretical prediction of mobility is not accurate even if bulk and interface electron-phonon couplings are active.

\subsection{Interplay between electron-phonon coupling and disorder strength}

In order to explain the experimental data, it is necessary to include also disorder effects. Indeed, there is evidence of traps in the bulk and at the interface with gates \cite{morpurgo}. Therefore, it is of paramount importance to investigate the role of disorder on the transport properties.

The model bears a strong resemblance with that of the previous subsections. Actually, the only modification is related to the electronic hamiltonian which includes here a disorder term. Therefore, in Eq. (\ref{hgen1}), there is a new term given by

\begin{equation}
H_{dis} = \sum_{i}  \epsilon_i n_{i} ,
\label{heldis}
\end{equation}
where $\epsilon_i$ is a local energy whose fluctuations in the range $[-W,W]$ simulate disorder effects in the bulk and at the interface with gate, $ n_i= c_{i}^{\dagger} c_{i}$ is the density operator, with $c_{i}^{\dagger}$ and $c_{i}$ electron creation and annihilation operators, respectively, relative to the site $R_i$. Due to the presence of shallow traps \cite{morpurgo}, disorder is not overwhelming and it is distributed according to a flat probability function.
The calculation method is analogous to that of the previous section.

\begin{figure}[htb]
\centering
\includegraphics[height=0.50\textwidth,angle=0]{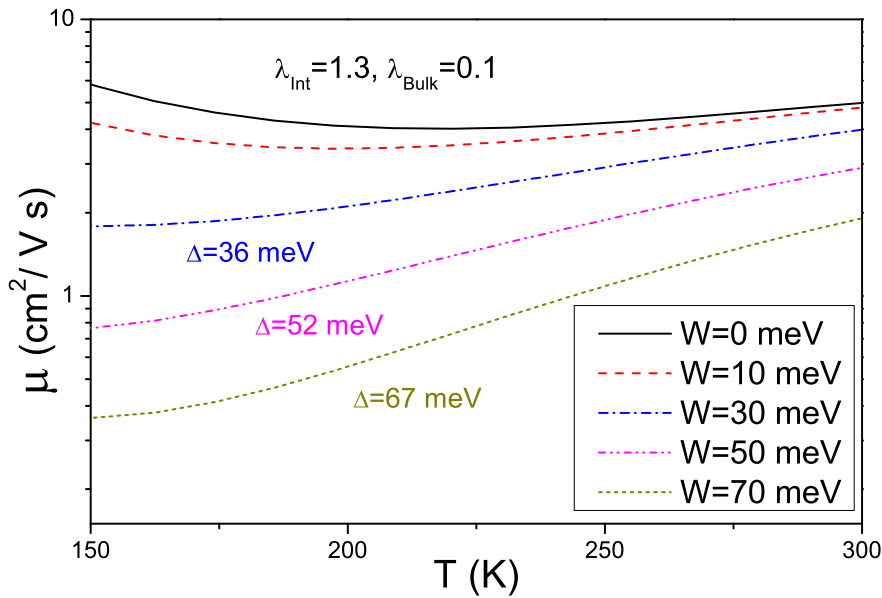}
\caption{Mobility $\mu$ as a function of the temperature for different disorder strengths $W$ at $\lambda_{Bulk}=0.1$ and $\lambda_{Int}=1.3$. The quantity $\Delta$ is the polaron activation energy.}
\label{fig7}
\end{figure}

In Fig. \ref{fig7}, we show the mobility as a function of the temperature with increasing the strength of disorder for $\lambda_{Int}=1.3$ and $\lambda_{Bulk}=0.1$. There are two main results. The first one is related to the suppression of the coherent metallic behavior with increasing $W$. The second one is the strong enhancement of the activation energy $\Delta$ up to $67$ meV even for the small amount of disorder $W=70$ meV. Furthermore, the decrease of the magnitude of the mobility is not so marked. Therefore, weak disorder effects are able to provide a very accurate description of the mobility resulting as key quantities for the interpretation of experimental data. Finally, another important effect of disorder is to drive the small polaron formation at lower electron-phonon couplings.


\section{Conclusions}

In this review, we have theoretically analyzed the effects of different electron-phonon couplings on spectral and transport properties of small molecule single-crystal organic semiconductors. Focus has been on oligoacenes, in particular on the series from naphthalene to rubrene and pentacene.

First, we have discussed the effects of the electron coupling to low frequency inter-molecular vibrational modes on the spectral and transport properties.
The resulting adiabatic models have been studied through numerical approaches with varying electron-phonon coupling and temperature. For rubrene, the model has considered the role of the electron-phonon coupling leading to a modulation of the particle hopping integral. With increasing temperature, the density of states is characterized by a larger exponential tail corresponding to localized states. Consequently, the chemical potential moves into the tail of the density of states, but this is not enough to drive the system into an insulating state. Not only the order of magnitude and the anisotropy ratio between different directions are accurate, but also the temperature dependence of the mobility is correctly reproduced in the model for rubrene. With increasing temperature, the Ioffe-Regel limit is reached since the contribution of itinerant states to the conduction becomes less and less relevant.

Then, we have analyzed the effects of electron coupling to both low frequency inter-molecular and high frequency intra-molecular modes on the spectral and transport properties. The interplay between local and non local electron-phonon interactions has been able to provide a very accurate description of the mobility of oligoacenes and to shed light on the intricate mechanism of band narrowing with increasing temperature. The band narrowing is a complicated phenomenon which could also be affected by the thermal expansion of the crystal structure \cite{libredas} (an effect which has not been analyzed in this review).

In the last part of the review, we have considered the influence of gates made of polarizable dielectrics on the transport properties. This effect has been
studied in a model which has combined bulk and long-range interface electron-phonon couplings. We have pointed out that the bulk coupling affects the behavior of mobility below room temperature enhancing the coherent contribution, but it is ineffective on the incoherent small polaron contribution dominated by the interface coupling at high temperatures.

Finally, we have emphasized the interplay between electron-phonon couplings and disorder strength on the transport properties. The presence of disorder is important to improve the modeling of the materials studied in this review. In particular, for systems gated with polarizable dielectrics, we have shown that disorder effects are able to enhance the hopping barriers of the activated mobility and to drive the small polaron formation at lower values of electron-phonon interactions. Therefore, disorder represents a key factor to get agreement with experimental data.

Some issues have not been covered in this review. Indeed, the transport properties could be affected by the nonlocal electron coupling not only to optical but also acoustic vibrations \cite{licorop}. The coupling to acoustic vibrations should be effective at low temperatures where it would be interesting also to investigate the role of quantum lattice fluctuations. These quantum effects are small in the adiabatic limit, however, they could be important in the regime where the presence of traps also influences the transport properties. Finally, we believe that concepts and methods discussed in this review can be a starting point for the study of related (such as durene crystals \cite{hanne4}) and more complex systems \cite{hanne5}.




\bibliographystyle{mdpi}
\makeatletter
\renewcommand\@biblabel[1]{#1. }
\makeatother


\end{document}